\documentclass[english]{article}
\usepackage[T1]{fontenc}
\usepackage[latin9]{inputenc}
\usepackage{amsmath}
\usepackage{amssymb}
\usepackage{wasysym}
\usepackage{esint}

\makeatletter
\usepackage{babel}

\makeatother

\usepackage{babel}
\begin{document}
\title{Hawking Radiation of Black Shells}
\author{Fernando Castro and J. Robel Arenas}
\date{Jul 2020}
\maketitle
\begin{abstract}
\noindent A black shell consists of a massive thin spherical shell
contracting toward its gravitational radius, coinciding with 't Hooft's brick wall, from the point of view
of an external observer far from the shell. This object was conceived
by J.R.Arenas and W.Israel in order to effectively model the gravitational
collapse dynamics and the thermodynamics of a black hole \cite{Arenas-Tejeiro,Castro,Pretorius,Israel66}.
We show in this article that a black shell presents the same emission
rate of a black hole when we consider Klein-Gordon equation in the
near horizon limit. We will use Parikh-Wilczek tunneling approach
to obtain the black shell emission rate \cite{Parikh-Wilczek}. 
\end{abstract}

\section{Introduction}

Bekenstein-Hawking entropy $S_{BH}$ has been derived from different
points of view \cite{Bekenstein,Mukohyama,Frolov}, and all agree
that it is proportional to surface area $A$ (in natural units):

\begin{equation}
S_{BH}=\frac{A}{4}\,.\label{C1.1}
\end{equation}

\noindent This suggests that the black hole thermal energy is strongly
concentrated near the horizon. With this idea, was conceived the black
shell model described above and the entanglement entropy of black
shells was obtained \cite{Arenas-Tejeiro}. Afterwards, physical entropy
of a black shell was derived from Gibbons-Hawking Euclidean approach
\cite{Castro}. 

\noindent In 1975, Hawking showed that black holes emit thermal radiation
with temperature (in natural units):

\begin{equation}
T_{H}=\frac{1}{4\pi R_{Sch}}.\label{C1.2}
\end{equation}

\noindent Where $R_{Sch}$ is Schwarzschild radius \cite{Hawking}.
In the same paper, Hawking gave an heuristic picture of the black
hole radiation as tunnelling of virtual particles across the horizon.

\noindent In this context Parikh and Wilczek presented a derivation
of black hole evaporation as a tunneling process \cite{Parikh-Wilczek}.

\noindent We show in this article that a black shell presents the
same emission rate of a black hole considering Parikh-Wilczek tunneling
approach \cite{Parikh-Wilczek}.

\section{Hawking Radiation of Black Shells}

In 1975 Stephen Hawking presented an heuristic image for the process
of radiation in terms of quantum tunnel effect of virtual particles
crossing the event horizon \cite{Hawking}. Subsequently, Parihk and
Wilczek presented a semiclassic derivation where the tunneling occur
near the horizon under energy conservation condition\cite{Parikh-Wilczek}.\bigskip{}

\noindent We consider in this section, a static thin shell of radius
$R_{0}$ outside and near the horizon according to Israel-Arenas black
shell model described above \cite{Arenas-Tejeiro,Castro}.

\noindent In this context, we employ the Darmois-Israel formalism
\cite{Darmois,Israel66,Musgrave} for a spherical thin shell $\Sigma$
which divides the space-time in two regions: the interior region $M^{-}$,
described by flat Minkowskian geometry and $M^{+}$, the exterior
geometry described by Schwarzschild spacetime. In what follows we
will use geometric units where: $C=G=\hbar=1$, and signature: $\left(-,+,+,+\right)$.

\noindent The exterior region $M^{+}$ is described by coordinates:
$X^{\alpha}=\left(t,r,\theta,\varphi\right)$, and the line element:

\begin{equation}
dS^{2}=-f\left(r\right)dt^{2}+f^{-1}\left(r\right)dr^{2}+r^{2}d\theta^{2}+r^{2}\sin^{2}\theta d\varphi^{2}.\label{C2.1}
\end{equation}

\medskip{}

\noindent Where: $f\left(r\right)=1-\frac{R_{sch}}{r}$ and $R_{sch}=2M$
is the Schwarzschild radius for a shell of mass $M$.

\noindent \medskip{}

\noindent The interior region $M^{-}$ is described by coordinates:
$Y^{\alpha}=\left(T,r,\theta,\varphi\right)$, and the line element:

\noindent 
\begin{equation}
dS^{2}=-dT^{2}+dr^{2}+r^{2}d\theta^{2}+r^{2}\sin^{2}\theta d\varphi^{2}.\label{C2.2}
\end{equation}

\noindent \medskip{}

\noindent The interior and exterior coordinates do not join smoothly
on the hypersurface $\sum$, but this doesn't matter because the junction
conditions are coordinate independent tensor equations \cite{Israel66}.

\noindent \medskip{}

\noindent The shell hypersurface $\Sigma$ is described using coordinates:
$\xi^{i}=\left(\tau,\theta,\varphi\right)$, and the 3-metric:

\noindent \medskip{}

\noindent 
\begin{equation}
dS^{2}=-d\tau^{2}+R_{0}^{2}\left(d\theta^{2}+\sin^{2}\theta d\varphi^{2}\right).\label{C2.3}
\end{equation}

\noindent \medskip{}

\noindent where $\tau$ is the proper time of the shell particles.

\noindent \medskip{}

\noindent Lets consider the massless Klein-Gordon equation for a scalar
field $\varphi_{g}$ , in the background metric $g_{\mu\nu}$ :

\begin{equation}
\frac{1}{\sqrt{\left|g\right|}}\partial_{\mu}\left(\sqrt{\left|g\right|}g^{^{\mu\nu}}\partial_{\nu}\right)\varphi_{g}=0.\label{C3.1}
\end{equation}

\noindent This equation can be written:

\begin{equation}
\frac{1}{\sqrt{\left|g\right|}}\partial_{\mu}\sqrt{\left|g\right|}g^{^{\mu\nu}}\partial_{\nu}\varphi_{g}+\partial_{\mu}\left(g^{^{\mu\nu}}\partial_{\nu}\right)\varphi_{g}=0.\label{8.1.1}
\end{equation}

\noindent Near the horizon and because tunneling occurs in radial
direction, the left term on the previous equation vanishes when $\omega\ll M$,
which gives:

\begin{equation}
\partial_{\mu}\left(g^{^{\mu\nu}}\partial_{\nu}\right)\varphi_{g}=0.\label{8.1.2}
\end{equation}

\noindent Expressing the field in terms of action $S$:

\begin{equation}
\varphi_{g}=e^{-\frac{i}{\hbar}S_{g}},\label{eq:8.2}
\end{equation}

\noindent and replacing in \eqref{8.1.2}, can be obtained:

\begin{equation}
\frac{i}{\hbar}\partial_{\mu}\left(g^{\mu\nu}\partial_{\nu}S_{g}\right)-\frac{1}{\hbar^{2}}\left(g^{\mu\nu}\partial_{\mu}S_{g}\partial_{\nu}S_{g}\right)=0.\label{eq:8.3}
\end{equation}

\noindent Solving for $\partial_{r}S_{g}$, in the real part of equation
\eqref{eq:8.3}, and because tunneling occurs in radial direction,
we obtain:

\begin{equation}
\partial_{r}S_{g}=\sqrt{-\frac{g^{tt}}{g^{rr}}}\omega.\label{eq:8.5}
\end{equation}

\noindent Being $\partial_{t}S=\omega$, the energy of the emitted
particle.

\noindent This solution also solves the complex part of the equation
\eqref{eq:8.3}, which can be verified by a direct calculation replacing
the solution \eqref{eq:8.5}:

\begin{equation}
\partial_{r}\left(\sqrt{-g^{rr}g^{tt}}\omega\right)=0\label{8.5.1}
\end{equation}

\noindent For a black shell, as previously stated, space-time is divided
into 2 regions:

\noindent \medskip{}

\noindent The outer region $M^{+}$ is described by Schwarzschild
space-time with metric $g$.

\noindent \medskip{}

\noindent The equation to be solved for $M^{+}$ and according to
previous considerations, is reduced to:

\begin{equation}
\APLbox_{g}\varphi_{g}=\partial_{\mu}\left(g^{^{\mu\nu}}\partial_{\nu}\right)\varphi_{g}=0.\label{eq:8.1-1}
\end{equation}

\noindent Whose solution is obtained by integrating the equation \eqref{eq:8.5}.

\noindent The inner region $M^{-}$ is described with Minkowskian
metric $\eta$.

\noindent \medskip{}

\noindent And the equation to be solved is:

\begin{equation}
\frac{1}{\sqrt{\left|\eta\right|}}\partial_{\mu}\left(\sqrt{\left|\eta\right|}\eta^{^{\mu\nu}}\partial_{\nu}\right)\varphi_{\eta}=0.\label{eq:8.1-1-1}
\end{equation}

\noindent Or equivalently:

\begin{equation}
\APLbox_{\eta}\varphi_{\eta}=\partial_{\mu}\left(\eta^{^{\mu\nu}}\partial_{\nu}\right)\varphi_{\eta}=0.\label{eq:8.1-1-1-1}
\end{equation}

\noindent The previous equation is linear, which allows to write it
as follows:

\begin{equation}
\partial_{\mu}\left(\left(\eta^{^{\mu\nu}}-g^{^{\mu\nu}}\right)\partial_{\nu}\right)\varphi_{\eta}+\partial_{\mu}\left(g^{^{\mu\nu}}\partial_{\nu}\right)\varphi_{\eta}=0.\label{eq:8.1-1-1-1-1}
\end{equation}

\noindent Or equivalently:

\noindent 
\begin{equation}
\left(\APLbox_{\eta-g}+\APLbox_{g}\right)\varphi_{\eta}=0.\label{eq:8.1-1-1-1-2}
\end{equation}

\noindent Is important to observe that the following equation is valid
regardless if $\eta^{\mu\nu}-g^{\mu\nu}$ are well defined metric
or solutions of Einstein field equations:

\begin{equation}
\APLbox_{\eta}=\APLbox_{\eta-g}+\APLbox_{g}\label{eq:18}
\end{equation}

\noindent Suppose that solutions of the previous equation, can be
written as: $\varphi_{\eta}=\varphi_{g}+\varphi_{\eta-g}+\varphi_{x}$,
then replacing it in \eqref{eq:8.1-1-1-1-2} shows that solutions
of $\varphi_{\eta}$ are linear combination of solutions of the following
equations:

\begin{equation}
\APLbox_{g}\varphi_{g}=0\label{eq:285}
\end{equation}

\begin{equation}
\APLbox_{\eta-g}\varphi_{\eta-g}=0\label{eq:286}
\end{equation}

\begin{equation}
\APLbox_{\eta}\varphi_{x}=-\APLbox_{\eta-g}\varphi_{g}-\APLbox_{g}\varphi_{\eta-g}\label{eq:287}
\end{equation}

\noindent The contribution to semiclassical emission rate of $\varphi_{\eta-g}$,
is obtained by solving the real part of :

\begin{equation}
S=-Im\left(\intop_{r_{i}}^{r_{f}}\sqrt{-\frac{\left(\eta^{tt}-g^{tt}\right)}{\left(\eta^{rr}-g^{rr}\right)}}\omega dr\right).\label{eq:8.7}
\end{equation}

\noindent Where: $\omega$ is the energy of the emited particle.

\noindent Replacing the inverse metric components:

\[
g^{tt}=-\frac{1}{f\left(r\right)},
\]

\[
g^{rr}=f\left(r\right),
\]

\[
\eta^{tt}=-1,
\]

\[
\eta^{rr}=1.
\]

\noindent Leads to:

\begin{equation}
S_{\eta-g}=\intop_{r_{i}}^{r_{f}}\sqrt{-\frac{\left(\eta^{tt}-g^{tt}\right)}{\left(\eta^{rr}-g^{rr}\right)}}\omega dr=\intop_{r_{i}}^{r_{f}}\sqrt{\frac{-r}{r-R_{Sch}}}\omega dr\label{eq:288}
\end{equation}

\noindent With the change of variable: $r=R_{Sch}+\epsilon e^{i\theta}$;
$dr=i\epsilon e^{i\theta}$, the solution is:

\begin{equation}
S_{\eta-g}=\underset{\epsilon\rightarrow0}{\lim}\intop_{0}^{\pi}-\sqrt{\epsilon}e^{\frac{i\theta}{2}}\sqrt{r_{p}+\epsilon e^{i\theta}}\omega d\theta=0\label{eq:289}
\end{equation}

\noindent This result shows that in the integration region $R_{Sch}-\epsilon\leq r\leq R_{Sch}+\epsilon$,
the solution $\varphi_{\eta-g}$ vanishes in the limit $\epsilon\rightarrow0$
:

\begin{equation}
\varphi_{\eta-g}=0.\label{eq:298}
\end{equation}

\noindent Replacing \eqref{eq:298} and \eqref{eq:8.5} in \eqref{eq:287}
shows that $\varphi_{x}=\varphi_{g}$ , and by this reason $\varphi_{g}$
is the only solution for the inner region $M^{-}$. We must note that
the interior and exterior coordinates do not join smoothly on $\Sigma$,
but they are both solutions of coordinate independent tensor equations.
Therefore, the junction conditions allow us to join the solutions
of both regions and say that $\varphi_{g}$ is solution for the entire
space: $M^{-}\cup M^{+}$. 

\pagebreak{}

\section{Conclusions}

\noindent Using the Parikh-Wilczek tunneling method \cite{Parikh-Wilczek}
and solving the Klein-Gordon equation in the near horizon limit, we
show that semiclassical emission rate of a black shell is the same
as that of a black hole. Israel junction conditions \cite{Israel66}
guarantees that this solution is valid for the entire space formed
by $M^{-}\cup M^{+},$ and we can conclude that Israel-Arenas black
shell emits thermal radiation with Hawking temperature.

\noindent The previous arguments could be extended to Reissner-Nordström
space-time.

\noindent \bigskip{}

\end{document}